\documentclass[10pt,conference]{IEEEtran}
\IEEEoverridecommandlockouts

\usepackage{cite}
\usepackage{amsmath,amssymb,amsfonts}
\usepackage{algorithmic}
\usepackage{graphicx}
\usepackage{textcomp}
\usepackage{xcolor}
\usepackage{hyperref}
\usepackage{tabularx}
\usepackage{threeparttable}
\usepackage{array}
\usepackage{ragged2e}
\usepackage{tablefootnote}

\def\BibTeX{{\rm B\kern-.05em{\sc i\kern-.025em b}\kern-.08em
    T\kern-.1667em\lower.7ex\hbox{E}\kern-.125emX}}
\begin{document}

\newcommand{\rep}{CG-Diff}
\newcommand{\sys}{Rhizome}
\newcommand{\todo}[1]{\textcolor{red}{[TODO: #1]}}
\newcommand{\dev}[1]{\textcolor{blue}{[Dev: #1]}}
\newcommand{\bimal}[1]{\textcolor{purple}{[Bimal: #1]}}

\DeclareRobustCommand*{\circled}[1]{\raisebox{-0.4ex}{\includegraphics[height=1.4\fontcharht\font`\B]{figs/circles/#1.pdf}}\hspace{0cm}}

\title{\rep{}: Organizing Code Changes Around Call Graphs}

\author{\IEEEauthorblockN{Bimal Raj Gyawali}
\IEEEauthorblockA{ UC San Diego \\
La Jolla, California, USA \\
bgyawali@ucsd.edu
}
\and
\IEEEauthorblockN{Devamardeep Hayatpur}
\IEEEauthorblockA{ UC San Diego \\
La Jolla, California, USA \\
dshayatpur@ucsd.edu
}
\and
\IEEEauthorblockN{Nadia Polikarpova}
\IEEEauthorblockA{ UC San Diego \\
La Jolla, California, USA \\
npolikarpova@ucsd.edu
}
}

\maketitle



\begin{abstract}

    Tools for code review present code changes file-by-file. We argue that, oftentimes, changes can be better organized around a call graph of the changed code.
    From an empirical study of GitHub pull requests (PRs) in-the-wild, we find that (1) in around 40\% of the PRs, more than half of the changed functions (and methods) are connected by call graphs, and (2) necessary callees are often located in different files from their callers. 
    Based on these findings, we develop the notion of \textit{CG-Diffs}, subgraphs derived by decomposing a call graph of the changed code into smaller and more navigable directed graphs.
    We then implement a web-based interface for viewing PRs that restructures the PR around these \rep{}s.
    Through a within-subject study comparing our interface against GitHub's PR view, we find that CG-Diffs help orient participants and provide them with more meaningful structures to navigate the PR.
    We also found several limitations: function nodes repeated across multiple CG-Diffs can be disorienting, and changes not contained in a function (e.g. globals, imports) are not as immediately apparent.
    Our study shows promise in using call graphs to help contextualize and navigate unfamiliar codebases, which may benefit new contributors to open source, and reviewers of unfamiliar LLM-generated PRs.
    
\end{abstract}
    


\begin{figure*}[t]
    \centering
    \includegraphics[width=\linewidth]{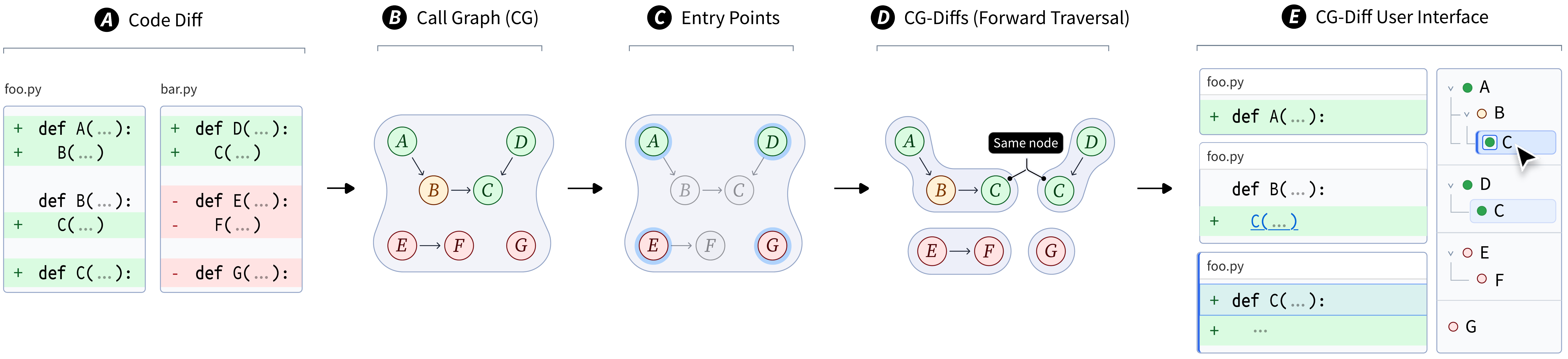}
    \caption{Construction of \rep{}s for an example PR 
    \circled{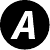}) The PR presented as a file-by-file diff, as shown in tools like GitHub.
    \circled{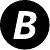}) A call graph is constructed from the changed functions. 
    \circled{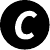}) Entry points, functions with no callers among the changed functions, are identified in colors.
    \circled{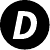}) The call chain is followed forward from each entry point; 
    \texttt{C} is reachable from both \texttt{A} and \texttt{D}. 
    \circled{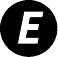}) The resulting \rep{}s are displayed in the user interface, showing the call tree and the code for the selected \rep{}.}
    \label{fig:cg-diff-construction}
\end{figure*}

\section{Introduction}

Code review is a widely adopted practice in software engineering where developers examine code 
written by peers before it is integrated into the codebase. It improves code quality, surfaces 
alternative solutions, facilitates knowledge transfer, and promotes shared ownership 
\cite{bacchelli2013expectations, yang2024survey, badampudi2023modern, sadowski2018google, macleod2017code}. 
In practice, code review takes place primarily through tools like GitHub and Gerrit.



These code review tools organize code changes file-by-file (e.g. \autoref{fig:cg-diff-construction} \circled{A}). However, file-based organization comes with challenges. It obscures how changes in one file \textit{relate} to another (a common question 
developers ask while reviewing code \cite{tao2012software}). Developers also pay less attention to files presented later in a diff \cite{fregnan2022first}, and in interviews, were either neutral or negative about the file-based reading order \cite{baum2017optimal}.

Prior work in program comprehension suggests that alternate organizations of source code, namely \textit{call graphs} \cite{ryder2006constructing}, can better facilitate code navigation and more quickly resolve developer queries \cite{latoza2010developers}. 
Based on an inkling that call graphs may also be a profitable representation for code review, we sought to answer the following:

\begin{enumerate}
    \item \textbf{RQ1:} How well-connected are real-world pull requests (PRs) by call graphs?
    \item \textbf{RQ2:} Do files split the callees necessary for understanding a caller?
\end{enumerate}

From a quantitative study of 9,310 PRs from 77 open-source projects, we found that in around 40\% of PRs, more than half of the changed functions and methods are connected by call graphs (\textbf{RQ1}). We then manually inspected 15 sampled PRs, finding that necessary callees are frequently in a different file from their caller (\textbf{RQ2}). 

These findings motivate the potential for structuring PRs via call graphs. However, naively presenting a call graph can be overwhelming and difficult to navigate\cite{herman2000graph}. We thus propose to construct smaller call graph diffs (``\rep{}s'') which are linear, directed, subgraphs of a call graph (\autoref{fig:cg-diff-construction} \circled{B} -- \circled{D}). We implement \rep{}s in a web-based user-interface for viewing PRs (\autoref{fig:cg-diff-construction} \circled{E}). We then conducted a within-subjects comparative study analysis comparing \rep{} against GitHub's PR view to investigate:

\begin{enumerate}
    \item[3)] \textbf{RQ3:} Does CG-Diff help developers navigate PRs?
\end{enumerate} 


We found that \rep{} helped participants get their bearings on the change before starting to read code in depth. \rep{} also provided structural cues that allowed participants to decide where to go next, and move between parts of the changed code more directly, relying less on searching or scrolling.

This paper makes the following contributions:

\begin{enumerate}

\item An empirical study quantifying how well-connected real-world PRs are by call graphs, and showing that necessary callees are frequently split across files from their callers.

\item \rep{}, a way of organizing PRs around call graphs, and a web-based interface for reviewing PRs using \rep{}.

\item A within-subjects user study showing that \rep{} helped participants orient within and navigate changed code more directly than GitHub's PR view.

\end{enumerate}

\section{Related Work}

\subsection{Cognitive Models of Programming and Code Review}
There are various cognitive models of how programmers read code. Brooks\cite{brooks1983towards} observed that developers form high-level hypotheses and then iteratively confirm and refine them. 
Koenemann and Robertson\cite{koenemann1991expert} found that developers forage cues as needed instead of following a systematic strategy. Rajlich and Wilde\cite{rajlich2002role} proposed that developers build understanding of code by mapping program elements onto domain concepts. These models vary, but they all share that developers recruit high-level conceptual strategies, and do not simply read code line-by-line. 

A smaller set of studies have described code review strategies. Gonçalves et al. \cite{gonccalves2025code} found that reviewers approach code review opportunistically. For small changes, reviewers read linearly. For large changes, they adopt various strategies: sometimes they address simpler parts of a change before moving to more difficult ones, and other times they begin with what they perceive as the core of the change. Gullstrand Heander et al.\cite{gullstrand2026code} proposed a two-phase model of code review with an orientation phase, and an analytical phase, where the reviewer first understands the rationale of a change, and then iteratively builds the understanding of the changed code. During these phases, reviewers require different kinds of information.

Tao et al. \cite{tao2012software} found that these information needs vary in difficulty: understanding the rationale behind a change is comparatively easy, while understanding how the change relates to other parts of the codebase is harder. They speculate that part of this difficulty comes from a  gap in tool support: static analysis tools operate on the whole codebase rather than at the level of a change, and code review tools fail to expose object-level relationships developers may need to effectively understand the change. Pascarella et al.\cite{pascarella2018information} found that reviewers ask questions to ensure they have captured the correct meaning of the code under review. These questions may be very specific (e.g., the behavior of a piece of code)  or generic (e.g., how that code fits into the broader system). They reported cases where some reviewers asked questions that could have been answered from the changed code, and were unable to relate a part of the change to others. Getting a response from the author for such questions can take several hours or even a day.

These studies paint code review as something reviewers work through in stages, needing different kinds of information at each one. Often, getting that information means looking past the file in front of them, tracing how a piece of code connects to others elsewhere in the change. However, existing tools do not explicitly support this relational view of code changes. We study one structure that captures this kind of connection, the call graph, to present the changed code.

\subsection{Tooling for Code Review}
The \textit{de facto} code review tools, like GitHub and Gerrit present code changes as a set of files, ordered alphabetically. Bouraffa et al. \cite{bouraffa2025not} found that nearly half of the reviews do not follow this default ordering. As the number of commented files increases, the order used by different reviews becomes closer to alphabetical, which they interpret as reviewers reverting to a default when no better strategy is available. Baum and Schneider \cite{baum2016need} observed that reviewers attempt to find an order that helps their understanding but often fall back to the order presented by their review tool. Bagirov et al. \cite{bagirov2023assessing} found that files appearing earlier in an ordering
receive more reviewer attention, and Fregnan et al. \cite{fregnan2022first} found that reviewers spend 
more time on the first file shown to them than on the last. Baum et al. \cite{baum2017optimal} ask reviewers for their preferred order,  and report that related changes should be placed close together, so that reviews can directly compare them and catch inconsistencies.
Söderberg et al. \cite{soderberg2022understanding} observed that developers look at changed code at different levels (e.g., checking coding convention vs. checking an API works correctly), which is not well supported by code review tools. Due to this, developers pull the changed code to their IDEs, which is time-consuming.

There are also various innovations atop ordinary code-review tools. ChangeViz \cite{gasparini2021changeviz} augments GitHub's interface with two lateral bars: a definitions bar showing where methods are declared, and a references bar showing where they are called, allowing reviewers to navigate between method calls and declarations. ReviewVis \cite{fregnan2023graph} uses a graph-based approach by visualizing classes and methods as nodes and their relationships, such as method calls, as edges.  In a study with developers, ReviewVis was perceived as helpful for mid-size changes, although as the change grew larger, the graph became harder to read \cite{fregnan2023graph}. ExplorViz \cite{krause2024visual} uses the three-dimensional software cities metaphor \cite{fittkau2013live} to visualize the changed code, representing packages as districts, classes as buildings, and calls between entities as streets whose thickness reflects call frequency. 

In each of these tools, the call graph is presented as a separate view that reviewers consult alongside the diff, whether as a sidebar, a graph, or a city view. The diff itself remains file-based, and reviewers use the call graph only as a reference while reading through files. We instead break the call graph into multiple linear, directed subgraphs and use each subgraph to lay out the diff: a caller, its callees, and their callees in turn are shown as blocks in call order (conceptually, each subgraph can be seen as a file).

\subsection{Call Graphs for Program Navigation}
Ko et al. \cite{ko2006exploratory} found that developers spend about a quarter of their time navigating code, either by following dependencies or by searching for names. To support developers in this process, researchers have proposed several systems that use call graphs.

Latoza et al. \cite{latoza2011visualizing} developed Reacher, which provides call graphs of the code for developers to navigate and helps them answer reachability questions \cite{latoza2010developers} across control flow paths.
Karrer et al. \cite{karrer2011stacksplorer} developed Stacksplorer, which adds two views alongside the code editor. The central source code editor displays the method selected by the developer. The left view shows methods calling the selected method, and the right view shows methods called from the selected method.
Krämer et al. \cite{kramer2012blaze} proposed Blaze, which supports two-phase navigation in the call graph. In the first phase, developers search for an anchor point they deem interesting. In the second phase, they follow paths starting from that anchor point until they reach the code they need. Blaze shows the path in the call graph to the method currently being edited, and allows developers to explore all paths involving that method.
Augustine et al. \cite{augustine2015field} developed Prodet, which uses developer’s recent activity to selectively visualize the call graph. It provides a map view to keep track of recently visited methods, and a search view for finding methods.

Although these tools vary in design, they all use call graphs to help developers navigate a codebase. Because real-world codebases are large and cannot be fully visualized, these tools focus on subgraphs around a method the developer selects. Our work differs in scope and intent. We operate on the changed code rather than the whole codebase, and use the call graph to determine how the diff is laid out.

\section{Empirical Study of PRs}
The thesis of this paper is that call graphs are a useful way to organize PRs. We thus first sought to understand how well-connected real-world PRs are by call-graphs (\textbf{RQ1}), and if files segments changes \textit{necessary} to understand a caller/callee relation (\textbf{RQ2}).

\subsection{How well connected are real-world PRs by call graphs?}
We sampled PRs from open-source Python projects on GitHub with 10k or more stars. We filtered for repositories with at least two contributors, excluding tutorials, course materials, and curated lists. This gave us 77 projects and 9,310 PRs. Of these, 1,820 contained no changed functions or methods, either because the changed files were not Python or changes occurred only outside functions. We report results for the remaining 7,490 PRs.

For each PR, we extracted the changed functions (and methods) and constructed a 
call graph. Some functions form clusters 
(groups connected by call relationships), and others are isolated, with 
no call relationships to other changed functions. To measure how connected a PR is, we define 
the \textit{Cluster Participation Ratio (CPR)} as the proportion of nodes that belong to 
a cluster. For example, in the PR from \autoref{fig:cg-diff-construction}, the call graph contains two clusters \{A, B, C, D\} and \{E, F\}, and one isolated node G (\autoref{fig:cg-diff-construction} \circled{B}), giving a CPR of $6/7\approx 0.86$.

\begin{figure}[t]
    \centering
    \includegraphics[width=0.9\linewidth]{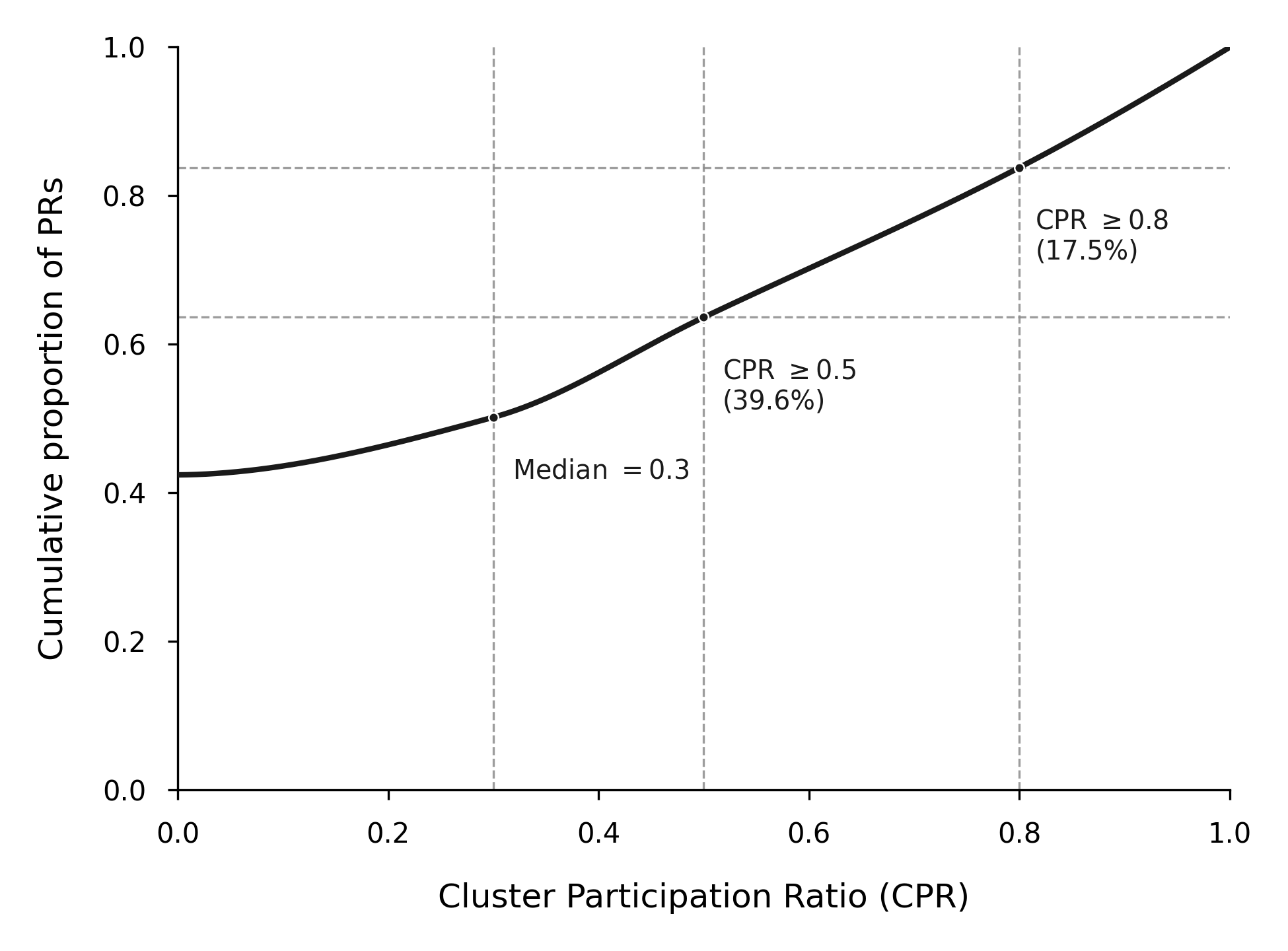}
    \caption{CDF of CPR for 7,490 PRs. 
In 39.6\% of PRs, more than half of the changed functions (and methods) participate in clusters (CPR $\geq$ 0.5). 
In 17.5\% of PRs, more than 80\% participate (CPR $\geq$ 0.8). The median CPR is 0.3.}
    \label{fig:cpr}
\end{figure}

\begin{figure}[t]
    \centering
    \includegraphics[width=0.9\linewidth]{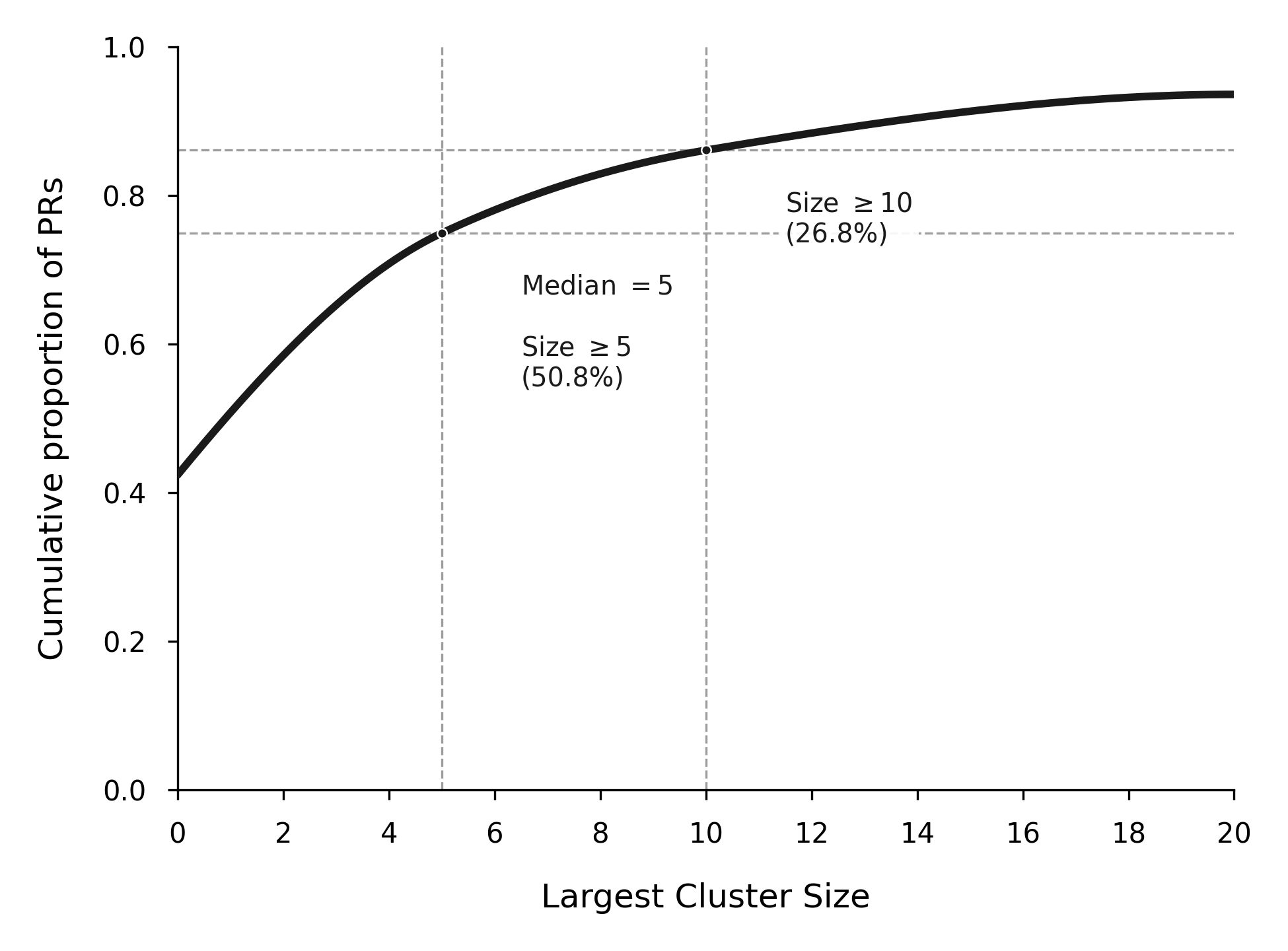}
    \caption{CDF of largest cluster size for 7,490 PRs. 
In 50.8\% of PRs, the largest cluster contains 5 or more functions (and methods). 
In 26.8\% it contains 10 or more. The median largest cluster size is 5.}
    \label{fig:size}
\end{figure}

Figure~\ref{fig:cpr} shows the distribution of CPR across all PRs. In 39.6\% of PRs more than 
half of the changed functions and methods participate in
clusters (CPR $\geq$ 0.5). The median CPR is 0.3.  In 50.8\% of PRs, at least one cluster contains 5 or more methods (Figure~\ref{fig:size}).

We observed that PRs vary widely in how connected they are by call graphs. In a typical PR, call graphs connect only 30\% of changed functions. Still, in nearly 40\% of PRs, more than half of them are connected by call graphs. This makes us optimistic that organizing PRs around call graphs could be useful for many code reviews. 


\subsection{Do files split the callees necessary for understanding a caller?}

The quantitative study for RQ1 showed that call graphs are present in a meaningful portion of 
real-world PRs. But from that study alone, we cannot tell whether the connected callees are necessary 
for a reviewer to understand the caller. A callee may be connected by a call graph but a reviewer might 
not need to read it to understand the caller (e.g., a helper function whose role is obvious from its name). Even 
when a callee is necessary, it may already be in the same file as the caller. Although files do not surface the 
call relationships, a reviewer might come across them more easily than if they were in a different file. Organizing 
PRs around call graphs would therefore be more valuable if callers and their necessary callees are located in 
different files.

We manually inspected 15 PRs with CPR between 0.5 and 0.9, where call graphs are 
sufficiently present while avoiding fully-connected PRs. We selected PRs with meaningful 
review activity (with at least 5 reviewer comments), filtering out those with similar call graph structure to already sampled ones 
to maximize variety. The first author read through each PR and coded caller-callee edge along 
two dimensions, for a total of 116 edges.

The first, \textit{call locality} (\texttt{D}), is whether the callee is in the same file as the 
caller (\texttt{D0}), a different file (\texttt{D1}), or outside the diff (\texttt{D2}).

The second, \textit{necessity} (\texttt{N}), is whether a reviewer needs to read the callee to understand 
the caller (\texttt{N1}) or not (\texttt{N0}). Examples of \texttt{N0} include 
helper functions whose purpose is evident from their name. The first author made this judgment, 
which we discuss as a threat to validity in \autoref{sec:threats}.

\begin{figure}[t]
    \centering
    \includegraphics[width=0.8\linewidth]{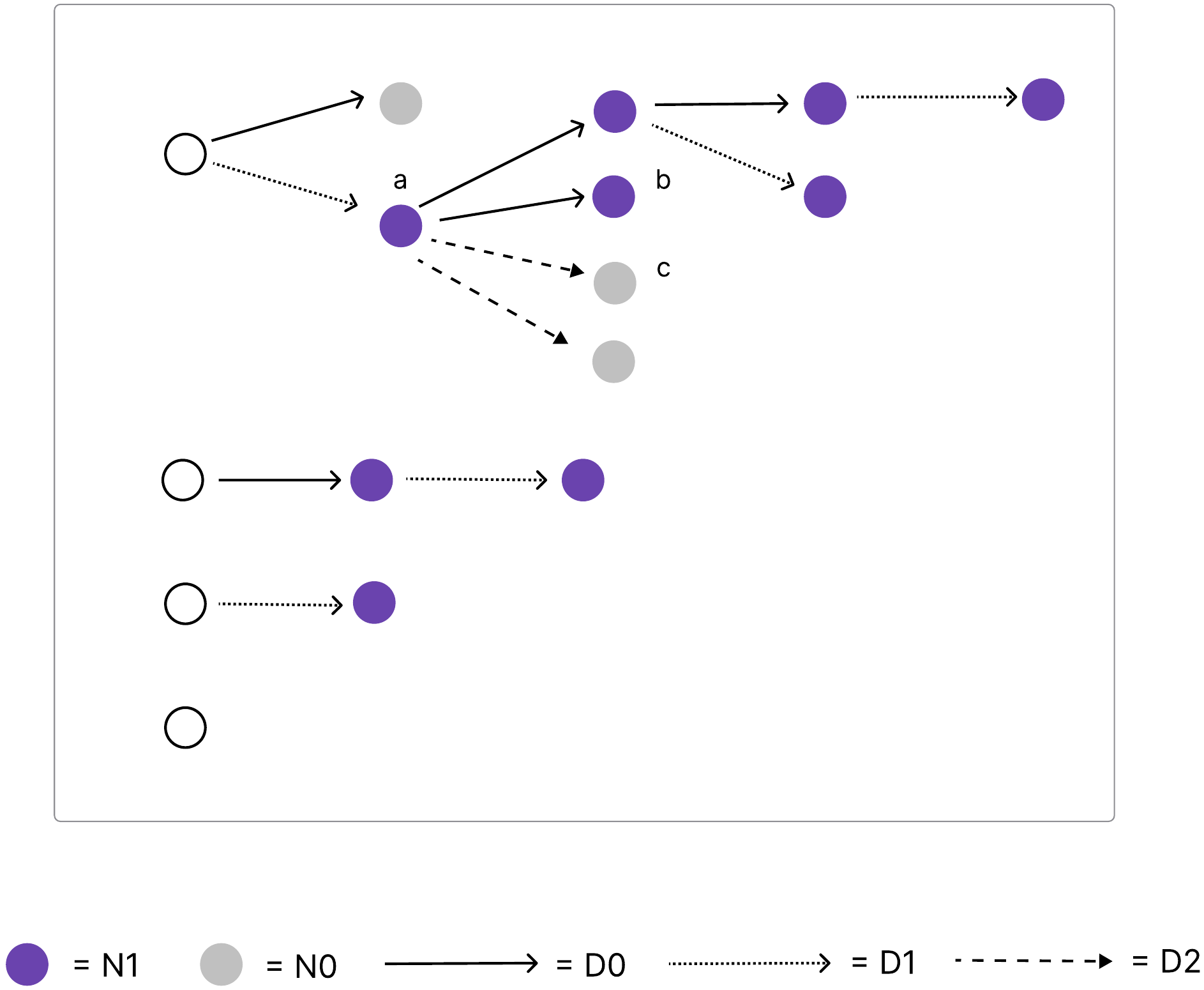}
    \caption{An example PR, coded along call locality (\texttt{D}) and necessity (\texttt{N}).
    Nodes are functions/methods, and edges are calls. Black nodes are entry points, 
    which are not callees of any other function and are therefore not coded. 
    Purple nodes are \texttt{N1} (necessary), grey nodes are \texttt{N0} (not necessary). 
    Solid edges are \texttt{D0} (same file), dotted edges are \texttt{D1} (different file), 
    dashed edges are \texttt{D2} (outside the diff).
    For example, function \texttt{a} calls \texttt{b} and \texttt{c}: \texttt{b} is necessary to understand \texttt{a}
    (\texttt{N1}) and in the same file as \texttt{a} (\texttt{D0}), while \texttt{c} is not 
    necessary (\texttt{N0}) to understand \texttt{a} and its definition is outside the diff (\texttt{D2}).}
    \label{fig:qual}
\end{figure}

Figure~\ref{fig:qual} shows an example PR\footnote{\url{https://github.com/huggingface/transformers/pull/41882}} from our sample with the two dimensions.

Across the 15 PRs, we found that 58 out of the 110 necessary callees were in a different file 
from their caller, and 52 were in the same file. 
The other 6 edges were coded as \texttt{N0} (not necessary).
None of the necessary callees were coded as \texttt{D2}, meaning all of them were contained within the diff.

Overall, a necessary callee in our sample was roughly as likely to 
be in a different file from the caller as in the same one.

\begin{figure*}[t]
    \centering
    \includegraphics[width=\textwidth]{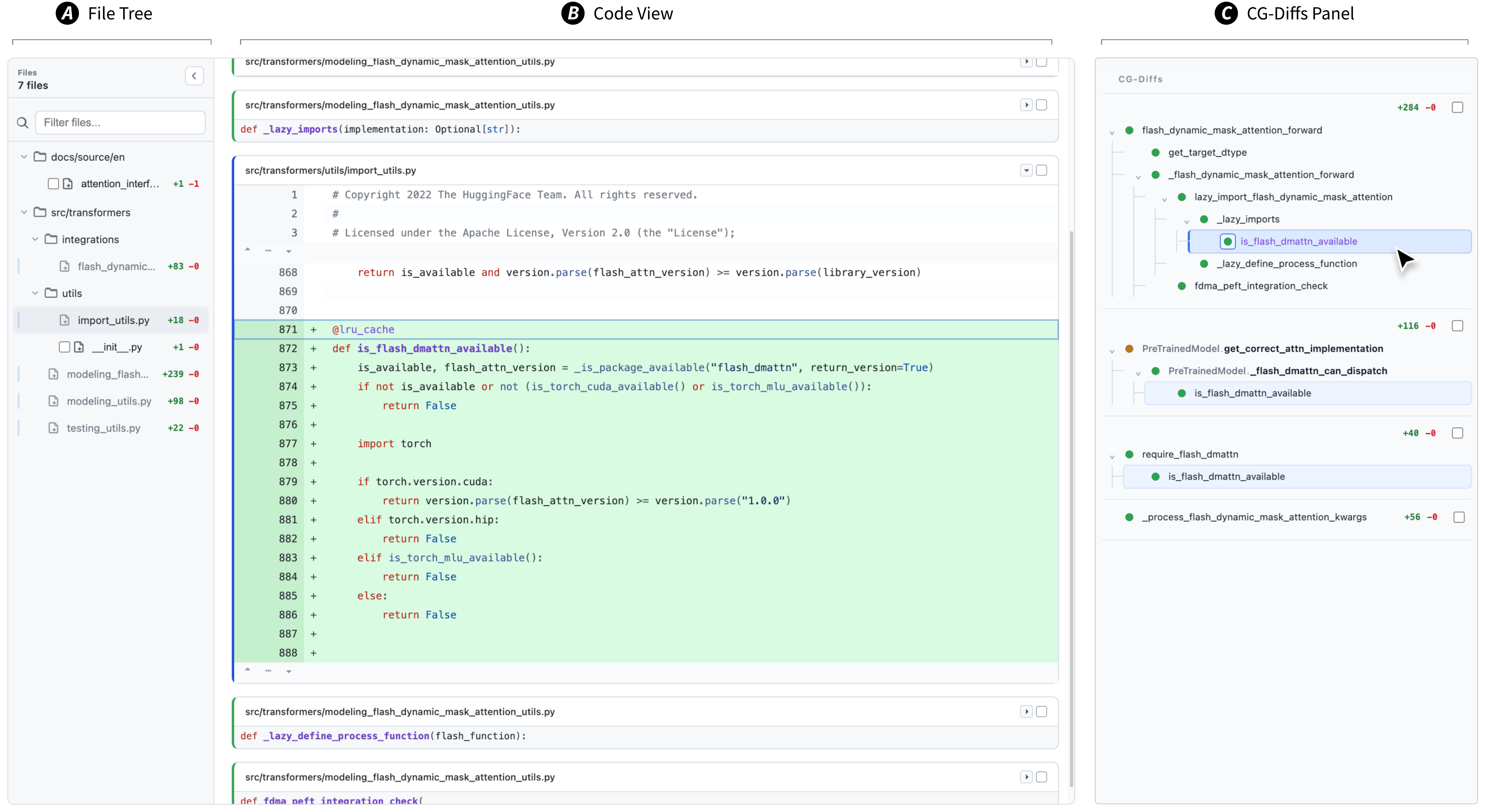}
    \caption{
        The \rep{} interface, laying out the PR as a set of \rep{}s.
        \circled{A}) The CG-Diffs panel showing four CG-Diffs; CG-Diff 1 is selected. 
        \circled{B}) The code view displaying only the functions/methods belonging to the selected CG-Diff. 
        \circled{C}) The file tree, showing changed files; files are not used for navigation.
    }
    \label{fig:tool_interface}
\end{figure*}

\section{\rep{}}
These findings supported our thesis that call graphs are a useful way to organize PRs, and motivated us to design \rep{},
which we describe in this section.

\subsection{Construction of \rep{}}

Naively presenting the full call graph of a PR can be overwhelming, as large graphs become difficult to read and navigate \cite{herman2000graph}. We instead break the call graph into smaller, linear subgraphs, which we call \rep{}s. We describe how we construct them below.

\begin{enumerate}
\item Consider a PR with changed functions \{\texttt{A, B, C, D, E, F, G}\}. We first construct a call 
graph of these functions, where edges represent calls between them (\autoref{fig:cg-diff-construction} \circled{B}).
\item Some functions have no callers (i.e. they have no incoming edges). In our example, these are \texttt{A, D, E}, 
and \texttt{G} (\autoref{fig:cg-diff-construction} \circled{C}). We refer to these as entry points, and follow the call chain forward.

\item  Traversing from \texttt{A} yields \texttt{\{A, B, C\}}; from \texttt{D} yields \texttt{\{D, C\}}; 
from \texttt{E} yields \texttt{\{E, F\}}; from \texttt{G} yields \texttt{\{G\}} (\autoref{fig:cg-diff-construction} \circled{D}). Each traversal produces a forward-directed subgraph rooted at the entry point, which we call a \rep{}. Since we do not revisit functions already encountered, each \rep{} is a directed acyclic graph.
\end{enumerate}

The PR is thus chunked into \rep{}s. Each starts with a single entry point, followed by its callees, and their callees in turn 
(\autoref{fig:cg-diff-construction} \circled{D}). Since the entry points have no callers among the changed 
functions, each \rep{} is self-contained. A function may appear in more than one \rep{} if it is reachable from 
multiple entry points; \texttt{C} appears in both the \texttt{A} rooted and \texttt{D}-rooted \rep{}s. An isolated 
function such as \texttt{G} yields a \rep{} of size 1, ensuring that no changed function is lost. 

\subsection{The \rep{} User Interface}

We use \rep{}s to lay out the PR. Think of a \rep{} as a container. Just like a file contains the functions written in it, a \rep{} 
contains the functions linked by a call graph. We lay out the PR as a set of \rep{}s instead of a set of files. We still show the file 
tree to provide additional context. \autoref{fig:tool_interface} shows the \rep{} interface 
for PR\footnote{\url{https://github.com/huggingface/transformers/pull/41882}}, 
which is composed of three linked 
views: \circled{A}) \textbf{\textit{File tree}}, \circled{B}) \textbf{\textit{Code view}}, and \circled{C}) the \textbf{\textit{\rep{}s Panel}}:

\subsubsection*{\textbf{File tree}} shows the changed files and highlights the file containing the currently selected function. For example, in \autoref{fig:tool_interface}, the file \texttt{import\_utils.py} is highlighted since it contains the selected function. Note that modified files cannot be directly used for navigation: they cannot be clicked and are only used to provide additional context. The primary way to navigate is by selecting a \rep{}. Files without any modified functions (i.e. do not participate in the call graphs) are clickable as usual to ensure that all parts of the changed code are included.

\subsubsection*{\textbf{Code view}} displays the functions of the selected \rep{} as a sequence of code blocks, each showing a full function definition with its source file indicated at the top. By default, all blocks are expanded and the reviewer can scroll through them as they would a normal diff. Function calls within each definition are rendered as links, allowing the reviewer to navigate directly to the called function. Each block can be collapsed or marked as read; since a function may appear in multiple \rep{}s, these states are reflected across all occurrences. In Figure~\ref{fig:tool_interface}, \texttt{is\_flash\_dmattn\_available} is shown expanded while the  remaining blocks are collapsed.

\subsubsection*{\textbf{\rep{}s panel}} presents the \rep{}s, 
each rendered as a tree rooted at its entry point, with callees nested beneath their caller in call order. Each \rep{} is annotated with its total number of changed lines, and \rep{}s are ordered by change size, with larger changes appearing first. 
A function may appear multiple times, either across \rep{}s, at multiple points within a single \rep{}, or due to recursion. In such cases, repeated occurrences are marked rather than expanded, avoiding infinite nesting. For example, in Figure~\ref{fig:tool_interface}, \texttt{is\_flash\_dmattn\_available} also appears in \rep{} 2 and \rep{} 3, where it is highlighted in those trees.
We add additional rules for \textit{class members} and \textit{global edits}:
\begin{enumerate}
    \item \textit{Class members.} 
 Sometimes, a \rep{} has a single node (i.e. no other changed functions share a call relationship), but it is still related to others by being a method of the same class. We therefore use class membership as an additional grouping signal by aggregating methods of the same class into a single \rep{}.
    \item \textit{Global edits.} A PR may contain changes outside function definitions, such as class definitions,  import statements, and global variables. Changed class definitions are shown as a class-based grouping (described above) if it exists, and otherwise attached to the \rep{}  containing the largest number of methods from that class. Changed import statements and global variables are attached to the function that references them most frequently and are displayed alongside that function's definition in the code view.
\end{enumerate}

\section{User Study of Developers using \rep{}}


To understand how CG-Diff helps developers navigate PRs (\textbf{RQ3}), we conducted a within-subjects qualitative study comparing
it against GitHub's PR view. We recruited seven developers with prior code review experience: three were professional 
software engineers and four were academic researchers (PhD/Postdoc). 
All participants were recruited through Slack groups and LinkedIn (Table~\ref{tab:participants}).

\begin{table*}[t]
\centering
\caption{Study Participants}
\renewcommand{\arraystretch}{1.1}
\begin{tabularx}{\linewidth}{l c X l X}
\hline
ID & Exp. & Title & Gender & Code Review Frequency \\
\hline
P1 & 1–3 yrs & Software Engineer & Male & Very frequently (multiple times a week) \\
P2 & 5+ yrs & Software Engineer & Male & Very frequently (multiple times a week) \\
P3 & 5+ yrs & Academic researcher (PhD/Postdoc) & Male & Frequently (weekly) \\
P4 & 3–5 yrs & Academic researcher (PhD/Postdoc) & Male & Very frequently (multiple times a week) \\
P5 & 1–3 yrs & Academic researcher (PhD/Postdoc) & Female & Occasionally (1–3 times a month) \\
P6 & 5+ yrs & Academic researcher (PhD/Postdoc) & Male & Occasionally (1–3 times a month) \\
P7 & 5+ yrs & Software Engineer & Female & Very frequently (multiple times a week) \\
\hline
\end{tabularx}
\label{tab:participants}
\end{table*}


Participants reviewed two PRs, one using GitHub and one using \rep{}, with both PR and tool order counterbalanced. To select the two PRs, we sampled PRs from our earlier analysis that were self-contained and comparable in complexity, i.e., with a similar number of \rep{}s and few references to code outside the PR. The first author manually inspected the PRs until six candidates were identified. Through pilot sessions with a separate set of participants, the first author selected two PRs for the study (Table~\ref{tab:PRs}).

Each session lasted approximately 90 minutes, and participants were compensated with a \$35 Amazon gift card for their time. The session consisted of the following:

\begin{enumerate}
    \item \textit{Introduction and Tool Walkthrough (10 minutes).} Participants were introduced to the study, with emphasis that the goal was to understand how they made sense of a PR rather than how many issues they could identify, and were asked to think aloud throughout. A guided walkthrough of CG-Diff using a separate demonstration PR was provided; all participants were already familiar with GitHub.
    \item \textit{Review First PR (25 minutes).} Before the review, participants received a brief description of the PR. They then reviewed the PR using the assigned tool while thinking aloud, describing what they were looking at, what they found important, and how they navigated.
    \item \textit{Follow-up Tasks (10 minutes).} After the review, participants were asked to: (1) summarize the PR, (2) describe how they would split it into three smaller commits if they were the author, and (3) describe the order in which they would review it again. These tasks were included to encourage participants to engage with the PR. The outcomes were comparable across conditions and are not analyzed further.
    \item \textit{Review Second PR (25 minutes).} Participants then reviewed a second PR using the other tool. The procedure was identical to the first PR review.
    \item \textit{Follow-up Tasks (10 minutes).} After completing the second PR review, participants completed the same follow-up tasks as in the first PR review.
    \item \textit{Semi-structured Interview (10 minutes).} After both PR reviews were completed, a semi-structured interview was conducted to elicit reflections on participants' experiences across the two tools. Participants were asked the following: \textit{(i) How did you feel your review experience differed across the two tools?}; \textit{(ii) Did you navigate the code differently between the two tools? If so, how did your navigation change?}; \textit{(iii) In each tool, how confident did you feel that you had seen most of the code changes?}
\end{enumerate}

The first author conducted thematic coding of participants' think-aloud PR reviews, video recordings, and semi-structured interviews, resulting in 85 initial codes (e.g., \textit{Difficulty in locating an important function}). These codes were discussed with the second author and iteratively refined and grouped into themes, which we present below.


\begin{table}[t]
    \centering
    \begin{threeparttable}
    \caption{PRs used in the user study}
    \renewcommand{\arraystretch}{1.3}
    \begin{tabularx}{\linewidth}{
        p{1.5cm}
    p{3.2cm}
    c
    c
    }
    \hline
    PR & Description & \# Functions & \# \rep{}s \\
    \hline
    \begin{tabular}[t]{@{}l@{}}
        Agno\\\#5709 \tnote{1}
    \end{tabular}
        & 
   Adds control over how and when database tables are set up across PostgreSQL, MySQL, and SQLite backends. & 28 & 9 
\\ 
    \begin{tabular}[t]{@{}l@{}}
        MarkItDown\\\#1263 \tnote{2}
        \end{tabular}
        & 
   Adds structured output instead of a single combined text block when converting documents to Markdown.
     & 10 & 9 \\
    \hline
    \end{tabularx}
    \label{tab:PRs}
    \begin{tablenotes}
        \fontsize{6pt}{7.5pt}\selectfont
        \item[1] \url{https://github.com/agno-agi/agno/pull/5709}
        \item[2] \url{https://github.com/microsoft/markitdown/pull/1263}
        \end{tablenotes}

        \smallskip
        \begin{quote}
        \footnotesize
        \textbf{Note:} 
        Both PRs were found comparable in review complexity by the first author and pilot participants.
        The Agno PR contains more changed functions because it includes additional helper functions.
        \end{quote}

        \end{threeparttable}
\end{table}

\subsection{\textbf{CG-Diffs structured initial exploration of the PR}}
In both the tools, participants started by getting a bearing on the change before investigating the code in depth. CG-Diffs 
helped orient participants during this process: P2 read through the tree top-down, building a sense for 
each part: \textit{``CG-Diffs basically tells you exactly what this is supposed to do. Like I think it's data 
tables. It first, I guess, created some namespace if needed. And then you build some metadata... it is kind of 
cool that you get this... [view of] basic blocks and how they're nested.''}
P2 then used CG-Diffs to conceptually chunk the PR: \textit{``one unit is clearly creating tables... this is how 
you do it for MySQL, this is how you do it for Postgres... and so there are basically two units.''}
Similarly, P3 found \textit{``[CG-Diffs] did give me a much better overall structure... I kind of at least figured
 out the intention [from looking at the CG-Diffs tree].''} P3, too, used the CG-Diffs tree to chunk parts of the 
 change:  \textit{``These changes are relevant to the definition of the page infrastructure and this one is about 
 the page converter... it's helpful for me to understand the overall code structure... they should belong to the 
 same type of changes.''}

With GitHub, participants were left to piece together a sense of the change from the list of files. P3 started by 
counting files and reading their names: \textit{``So there are 7 files. 1, 2, 3, 4, 5, 6, 7. I understand each of 
them [inferred from the file names]... So I will just start from the first.''} Participants also used other starting 
points, like P4, who \textit{``will definitely review the test files as a first file because we don't have any main 
function here. We have just a set of functions.''} P2 reflected that these strategies might not always be effective, especially 
when they did not already know the codebase:  \textit{``with just a plain GitHub, you kind of have to go down the list of files... if you 
know the code base... you start from here and go because you know exactly where things are. But if you don't, you have no choice.''}

 \subsection{\textbf{CG-Diffs provided an overview of the PR}} 
 In CG-Diff, the changes were co-located in an easily legible representation. P5 described using it as a 
 lookup: \textit{``So for now, I'm using the CG-Diffs as a table of contents, actually.``} Seeing everything at 
 once, some participants found themselves noticing things and drawing connections across the change. P3, for 
 example, noticed functions they considered important: \textit{``there is a separate main function... You can 
 easily notice like this one... this one defines the overall API.''}
In GitHub, this single place did not exist. P5 described the difficulty: \textit{``I don't have a center of all new changes for me to navigate the PR and that's... the workload is quite high to start understanding what is changed versus not.``}

    \subsection{\textbf{Structural cues in CG-Diffs helped participants navigate}}
    Once participants had an initial sense of the change, they moved to a detailed inspection of the code. 
    Structural cues in CG-Diffs enabled participants to glance at the tree and infer their relevance, 
    before committing to a closer read. 
    
    P2, for example, saw a base class and its concrete implementations (e.g., for MySQL) 
    together in the CG-Diffs tree, and used this to guess that the base class was a shared interface rather than where the 
    logic they needed actually lived, before moving on to read the MySQL implementation more closely.
    Similarly, P7 saw in the CG-Diffs tree that each backend implementation included the same 
    function, and used this to decide which to inspect individually: \textit{``Okay, so all of these different engines have 
    a create all tables. So now I'll go individually in this MySQL.''} 

    When using GitHub, participants relied on file and function names as signals for relevance and often ended up reading 
    the code more fully just to figure out whether it was relevant. 
    For example, P4 spent some time reading a function they initially expected to be the one they needed, 
    only to later realize otherwise: \textit{``I suppose this function is our function of interest. Actually, no.''}

   \subsection{\textbf{CG-Diffs allowed participants to be agile}} 
    During review, participants raised questions that required them to move between different parts of the 
    change, such as where a function was defined or which other functions called it. In CG-Diffs, 
    participants could click directly to what they needed. P1, for example, used the right panel 
    to navigate to a function's definition and see the changed code: \textit{``if you see on 
    the right side you have all the functions definitions described. So if you click on the function's 
    definition then you can see whatever the code that has been different. I can easily navigate.''} 
    P4 similarly described moving across the call chain: \textit{``that was a pretty 
    good feature of CG-Diffs that you can click on a function and see it... back to the 
    place where it was called... and go to all of the invocation sites.''} 

    In GitHub, answering these same questions meant going looking for the answer themselves, 
    with whatever tools were at hand. P3, for example, searched by hand for where a function was 
    called: \textit{``I remember I manually searched where each part of these functions are called anywhere to 
    understand their relations.''} P4 turned to the browser's built-in search as a stand-in: \textit{``on GitHub, 
    I still cannot review PRs without this overkiller, which is Firefox search feature.''} This was not 
    always effective: P1 spent considerable time searching for a function and gave up without finding 
    it: \textit{``I'm just looking for this document converter result. I'm not able to find it... I'm still not 
    able to find where this method has been written though.''}


\subsection{\textbf{Participants started from the file, even when reviewing through CG-Diffs}}
Some participants (P4, P6) initially interpreted CG-Diffs through the file-based view, anchoring themselves in the 
file-based view before engaging with the new representation. For example, P6 deliberately chose to begin with 
the file view before engaging with CG-Diffs: \textit{``I'm just going to look at the diff from the files on the 
left-hand side, and not try and let the right-hand side affect me too much until, I can figure out how it can be 
useful.''} Similarly, P4 started by locating a file in the file tree and then mapped it to its corresponding 
CG-Diff in the CG-Diffs tree: \textit{``I need to look here [in the file tree], and I need to find this file [in CG-Diffs tree], if I
want this file, or that file, I want to start with it.''}

\subsection{\textbf{Initial friction with CG-Diffs eased with use, and participants came to see it as more effective}}

P4, who started by anchoring in the file view, reflected that with more use, CG-Diffs would become more efficient 
than reviewing code without any structure: \textit{``Compared to GitHub, I really like CG-Diffs and I would use this instead of 
GitHub to review my PRs... I think if I would use CG-Diffs for a really long period of time I would understand how to use it 
more efficiently.''} P2 similarly attributed the early friction to simply being new to the tool: \textit{``As with 
any new tool, it's kind of, you see something new and then there are all these kind of distractions...I've used GitHub 
my whole life, if I had CG-Diffs for a week, maybe the experience would be completely different. I did like the option
 to view change this way.''} P2 went on to note that the structure CG-Diffs surfaced was already familiar in another sense: 
 \textit{``When I have to do review unfamiliar PRs, I open them in my IDE so I can exactly follow the references, 
 and kind of build those CG-Diffs in my mind, and so here this was useful.''}

\subsection{Limitations of \rep{}}

We found some limitations arising from the design of \rep{}, which we describe below.

\paragraph*{\textbf{Repetition of functions across CG-Diffs caused confusion when the PR itself also contained repetitive code}}
Functions reachable from multiple entry points appear in 
multiple CG-Diffs, so participants sometimes found it difficult to differentiate 
whether repetition came from the PR or from how \rep{} presented it. 
P6 described
the difficulty: \textit{``CG-Diffs shows repeated blocks of code and the code itself is repetitive. So that mulled up my brain because 
I was not sure where the repetition was coming from.''} P6 described a case: \textit{``there was a function that was only 
defined once. But because I saw it multiple times [repeated by CG-Diffs], I thought it was also repeated [in the PR]. But then the other function 
was repeated across all the different backends [actually repeated in the PR itself]...''} More generally: \textit{``It's easy to get 
overwhelmed... it's hard to keep track of which functions are the same.''}

\paragraph*{\textbf{Placement of code outside functions was not always apparent}}
The code outside functions (e.g., globals, imports) was placed in the CG-Diff that references it most, 
but this was not always apparent to participants, who sometimes felt that changes were missing from the view. P7 felt that some changes did not 
appear in the CG-Diffs: \textit{``I feel like there were some changes that are not mentioned in the CG-Diff, but that did not hamper 
my understanding of the PR. I found this PR through this process much easier to understand than just GitHub.''}

\section{Discussion and Future Work}

In our study, we found that \rep{} helped participants get an overview of the PR and navigate between its parts, relying less on searching or scrolling. Below, we discuss the broader implications of organizing code changes around call graphs.

\subsection{CG-Diffs can be particularly valuable in unfamiliar codebases} 

Code review tools assume some degree of familiarity with the codebase: the file-based organization is only navigable if the reviewers already know the meaning of each file and how they relate to each other.

Our study showed that CG-Diffs can be effective at surfacing relationships and structure in an \textit{unfamiliar} codebase. Moreover, prior work in codebase onboarding suggests that a structural overview, i.e., a rough outline of the shape of the code and how parts relate, like CG-Diffs, is especially valuable 
to developers working in unfamiliar codebases \cite{yates2020characterizing}. As such, we envision that the greatest benefit of CG-Diffs will be felt by reviewers who are less familiar with the codebase or first time contributors. Recent advances in LLM-assisted development are also often structured around `agents' generating new PRs for the programmer (who often takes a less hands-on role) to review -- another workflow that can potentially benefit from CG-Diff's structured representations.

\subsection{Studying CG-Diffs in personally meaningful codebases and over time}
Reviewers approach a PR in different ways, shaped by their familiarity with the codebase, the nature of the change, and 
their own reviewing habits \cite{gonccalves2025code}. In our study, all participants were equally new to both CG-Diffs and 
the repositories they reviewed. Therefore, the study describes how CG-Diffs helps a reviewer build understanding \textit{de novo}, but does not characterize how CG-Diffs fits into the workflow of a user already familiar with a codebase.

Even so, we observed differences in how participants' navigated the PR. Some used CG-Diffs as a table of contents to opportunistically move through the changed code (as one may expect from someone who is more familiar with the codebase). Others moved through the PR linearly: CG-Diff by CG-Diff. 

A natural question to ask is whether CG-Diffs' role changes when reviewing a codebase one is already familiar with. A reviewer who knows a codebase well may find less value in CG-Diffs, or may use it differently (e.g., as a sanity check rather than a representation they frequently interact with). A more longitudinal study which follows developers' use of CG-Diffs over time can shed light on this question. We suspect that the guiding thesis of CG-Diff, i.e., organizing the change around call graphs, will always remain useful to consult -- though the form of the use may change.

\section{Threats to Validity}
\label{sec:threats}

\textbf{\textit{Participant sample.}} With seven participants, we do not claim 
our findings generalize to every developer or every codebase. 
We observed convergence across participants, and across two PRs which gives us confidence 
in the themes we report.
Still, seven participants is a small window into the much larger range of experience, habits, and review 
practices that exist among developers, and we cannot rule out that a different sample would surface different or 
additional themes.

\textbf{\textit{Unfamiliar codebases.}} Code review in practice usually takes place in a codebase the reviewer has some 
familiarity with. Our participants, 
by contrast, were seeing each codebase for the first time, and while we gave them a brief description of the PR beforehand, 
this does not replicate the working knowledge a reviewer builds up over time on a project they maintain. 
Our findings may therefore be more representative of reviewing unfamiliar code than of code review more broadly.

\textbf{\textit{Subjective coding.}} Judging the necessity (\texttt{N}) of a function in our manual inspection of 15 PRs relied on the first author's understanding 
of the PR. Another reviewer might reasonably judge the necessity of some functions differently, introducing subjectivity
into the coding used for RQ2.
\section{Conclusion}
File-by-file organization has been the default for code review tools for decades, and its limitations have been
 documented but largely accepted. This paper asked whether code changes can be organized around
 call graphs instead. The answer, in the contexts we studied, is encouraging.

 We motivated our work with quantitative study of 9,310 PRs in the wild, and manual inspection of 15 sampled PRs, 
 finding that call graphs are sufficiently present in a meaningful fraction of PRs, 
 and that necessary callees to understand a caller are often split by files.
 We then proposed CG-Diffs, which are linear, directed subgraphs of a call graph of the changed code. Through a 
 user study with developers comparing CG Diffs view with GitHub PR view, we found that CG-Diffs provided 
 meaningful structure that helped participants orient and navigate the PR.
 Our results suggest that call graphs can be a useful way to present code changes in review tools.

\bibliographystyle{IEEEtran} 
\bibliography{IEEEabrv,references}

\end{document}